\documentstyle[amsfonts,epsfig,12pt]{article}

\newcommand {\eqref} [1] {(\ref {#1})}

\newcommand {\slsh} [1] {\not{\hbox{\kern-2pt${#1}$}}}

\def\drawbox#1#2{\hrule height#2pt
         \hbox{\vrule width#2pt height#1pt \kern#1pt
               \vrule width#2pt}
               \hrule height#2pt}

\def\Asym#1#2{\vcenter{\vbox{\drawbox{#1}{#2}
               \kern-#2pt       
               \drawbox{#1}{#2}}}}

\newcommand {\beq} {\begin{equation}}
\newcommand {\eeq} {\end{equation}}
  \newcommand {\ber}{\begin{eqnarray*}}
  \newcommand {\eer} {\end{eqnarray*}}
\newcommand {\bea}{\begin{eqnarray}}
  \newcommand {\eea} {\end{eqnarray}}

\newcommand{\Dslash}{\,{\raise.15ex\hbox{/}\mkern-12mu D}}

\begin{document}


\begin{titlepage}

\begin{center}
\vspace{1in}
\large{\bf Noncommutative Two-Dimensional Gauge Theories}\\
\vspace{0.4in}
\large{Adi Armoni}\\
\small{\texttt{a.armoni@swan.ac.uk}}\\
\vspace{0.2in}
\large{{\emph Department of Physics, Swansea University}\\ 
\emph{Singleton Park, Swansea, SA2 8PP, UK}\\}
\vspace{0.3in}
\end{center}

\abstract{We elaborate on the dynamics of noncommutative two-dimensional gauge field theories. We consider $U(N)$ gauge theories with fermions in either the fundamental or the adjoint representation. Noncommutativity leads to a rather non-trivial dependence on theta (the noncommutativity parameter) and to a rich dynamics. In particular the mass spectrum of the noncommutative $U(1)$ theory with adjoint matter is similar to that of ordinary (commutative) two-dimensional large-$N$ $SU(N)$ gauge theory with adjoint matter. The noncommutative version of the 't Hooft model receives a non-trivial contribution to the vacuum polarization starting from three-loops order. As a result the mass spectrum of the noncommutative theory is expected to be different from that of the commutative theory.}

\end{titlepage}

\section{Introduction}

Noncommutative gauge field theories exhibit fascinating dynamics due to the so-called ``UV/IR mixing'' phenomenon. It was shown \cite{Minwalla:1999px,Matusis:2000jf} that short distance effects, which would naively be considered as irrelevant, could alter the infra-red dynamics. Surprisingly, an infinite sum of naively irrelevant operators conspire to become relevant.

Let us demonstrate the above statement in a rather simple example: the non-planar contribution to the trace of the vacuum polarization in a four dimensional pure $U(1)$ gauge theory \cite{Matusis:2000jf}

\beq
\Pi  _{\rm non-planar} (p) = \int {d^4 k \over k^2} \exp i p \theta k
 = -{1 \over (\theta p)^2} \, . \label{fourd}
\eeq   

Had we expanded the exponent of \eqref{fourd} we would have obtained
\beq
 \sum _{n=0}^{\infty} \int {d^4 k \over k^2} {(i p \theta k)^n \over n!} =  \sum _{n=0}^{\infty} {(i \theta p)^n \over n!} \Lambda ^{(n+2)}
\eeq
with $\Lambda$ a UV cut-off. Namely, instead of a dimension two operator \eqref{fourd}, we would obtain an infinite sum of irrelevant operators. 

It turns out that not only the effective theory acquires a dimension two operator, but in fact in the pure YM case the photon becomes a tachyon. Gauge invariance does not protect the photon from acquiring a mass. Due to the explicit breaking of Lorentz invariance the gauge theory becomes perturbatively unstable. It is not clear whether it admits a stable vacuum and the issue is rather similar to the question of whether bosonic string theory admits a vacuum \cite{Armoni:2001uw}.   

The situation in two-dimensional gauge theories is simpler. Noncommutativity in two dimensions does not break Lorentz invariance, since in two-dimensions
\beq
[x^\mu, x^\nu] = \theta ^{\mu \nu} = \theta \epsilon ^{\mu \nu} \,,
\eeq
hence gauge bosons cannot acquire a mass \footnote{Except the standard 2d Schwinger mass \cite{Schwinger:1962tp} which should better be understood as a mass for the meson of the theory.}. Moreover, since the dynamics of field theories in two dimensions is much simpler than the dynamics of four dimensional theories, two-dimensional theories are a good starting point for the understanding of how noncommutativity affects the dynamics. The two-dimensional noncommutative Schwinger was recently analyzed in \cite{Ardalan:2010qb}, where it was found that at the one-loop level the mass spectrum of the commutative and the noncommutative theories is the same.

We will consider noncommutative two-dimensional $U(N)$ theories with matter in either the adjoint representation or in the fundamental representation. Our aim is to examine how noncommutativity affects the two-dimensional dynamics. We will find that noncommutativity could make drastic changes. In particular, whereas the commutative $U(1)$ theory coupled to an adjoint fermion is a free theory, the noncommutative version of the theory is a highly non-trivial theory which, presumably, admits an infinite set of parallel Regge trajectories.

The paper is organized as follows: in section 2 we discuss the general structure of two-dimensional noncommutative theories. In section 3 we consider a $U(N)$ gauge theory with matter in the adjoint representation. In section 4 we comment on the noncommutative version of the 't Hooft model. Section 5 is devoted to a discussion.

\section{UV/IR mixing in two dimensions}

Commutative two-dimensional are super-renormalizable, as the gauge coupling has a dimension of mass. UV divergences may occur only at the one-loop level. Higher loops are finite. 

Let us consider the noncommutative version of the theory and focus on the gluon vacuum polarization. Since in two-dimension noncommutativity does not break
Lorentz invariance or gauge invariance, a general expression for the vacuum polarization is
\beq
\Pi ^{\mu \nu}(q)=(q^2 g^{\mu \nu} - q^\mu q^\nu)\Pi(q^2,e^2, \theta) \,.
\eeq
Since noncommutative effects arise from non-planar graphs it is useful
to decompose $\Pi$ into planar and non-planar contributions
\beq
\Pi \equiv \Pi _{\rm planar} (q^2, e^2) + \Pi_{\rm non-planar} (q^2,e^2, \theta) \,.
\eeq

Planar graphs are identical to the corresponding graphs in the commutative theory, so they do not depend on $\theta$. Since $\Pi$ is dimensionless, the perturbative expansion is parameterized by $e^2 / q^2$. 

 Non-planar graphs in noncommutative theories differ from the corresponding
graphs in commutative theories. Let us analyze the general behavior of
a non-planar graph in a noncommutative 2d theory. At loop order $l$ it is
\beq
\Pi_{\rm non-planar} (q^2,e^2, \theta)\sim {(e^2)^l \over q^2} \int  dt t^{(l-2)} \exp \left ( -{(\theta q)^2 \over t} -t(q^2+m^2) \right ) \,. \label{general} 
\eeq
The above expression \eqref{general} represents a schematic form of $\Pi_{\rm non-planar}$ (the actual $l$-loop amplitude is written in terms of $l$ Schwinger parameters, not one). Specific examples will be discussed in the following sections. Few comments are in order: one power of $1/q^2$, in front of the integral, is due to gauge invariance, namely due to the form of $\Pi ^{\mu \nu}$. The rest follows from dimensional analysis and general properties of Feynman graphs, as discussed in the appendix of ref. \cite{Minwalla:1999px}. Small values of $t$ correspond to the IR, while large values of $t$ correspond to the UV. When $\theta=0$ the above integral diverges logarithmically at small $t$ only at the one-loop order ($l=1$). It converges for $l>1$, due to the fact that the theory is super-renormalizable. Since that integral is finite for $l>1$ even for $\theta=0$, it means that for non-zero $\theta$, we can expand in powers of $\theta$ and the limit $\theta \rightarrow 0$ is smooth. The case $l=1$ is more subtle and as we shall see in the next section, the limit $\theta \rightarrow 0$ becomes singular when there exists a non-planar one-loop contribution to the vacuum polarization.

\section{Theories with adjoint matter}

Consider a 2d noncommutative $U(N)$ gauge theory coupled to one-flavor
of a Dirac fermion that transforms in the adjoint representation of the gauge group. The action of the theory is
\beq
S=\int d^2 x \, {\rm tr} \left ( -{1\over 2e^2} F_{\mu \nu} \star F ^{\mu \nu} + \bar \Psi i \slsh \partial \star \Psi + A_\mu \star \bar \Psi \gamma ^\mu \star \Psi - \bar \Psi \star A_\mu \gamma ^\mu \star \Psi \right )\, ,
\eeq 
where both the gauge field and the fermion are $N \times N$ matrices. $F_{\mu \nu}= \partial _\mu A_\nu - \partial _\nu A_\mu - i(A_\mu \star A_\nu - A_\nu \star A_\mu)$. The $\star$-product is defined as $f\star g(x)= \exp ({i\over 2} \theta ^{\mu \nu} \partial ^\eta _\mu \partial ^\xi _\nu) f(x+\eta)g(y+\xi)|_{\eta,\xi \rightarrow 0}$.

Let us consider the contributions to the vacuum polarization\footnote{Note that in 2d the only contribution at one-loop order is from a fermionic loop, since gluons carry $(d-2)$ degrees of freedom. The easiest way to understand it is in the chiral gauge where the Yang-Mills Lagrangian is ${\cal L}={\rm tr}\, (\partial _- A_+)^2$}. The planar contribution is as in the commutative theory
\beq
i\Pi ^{\mu \nu} _{\rm planar} = -e^2N \int {d^2 l \over (2\pi)^2} {\rm tr} 
\left ( \gamma ^\mu {\slsh l \over l^2} \gamma ^\nu {(\slsh l + \slsh q) \over (l+q)^2 }\right )  \label{planar}
 \eeq
The non-planar graph is depicted in figure \eqref{graphs2} below. It is non-vanishing only when the external legs are $U(1)$ gauge bosons \cite{Armoni:2000xr}. The expression is
\beq
i\Pi ^{\mu \nu} _{\rm non-planar} = e^2 \int {d^2 l \over (2\pi)^2} {\rm tr} 
\left ( \gamma ^\mu {\slsh l \over l^2} \gamma ^\nu {(\slsh l + \slsh q) \over (l+q)^2 } \exp i l \theta q \right )  \label{nonplanar}
 \eeq

\begin{figure}[!ht]
\centerline{\includegraphics[width=4cm]{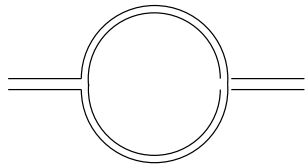}}
\caption{\footnotesize A non-planar one-loop contribution to the photon vacuum polarization.} \label{graphs2}
\end{figure}

Let us focus on the $U(1)$ theory. When $\theta =0$ it is clear that
\beq
\Pi ^{\mu \nu} _{\rm planar}+\Pi ^{\mu \nu} _{\rm non-planar} =0 . \label{zerosum}
\eeq
The above result makes perfect sense: in the commutative case the adjoint fermion decouples from the photon and the theory is free.

Let us consider the noncommutative $U(1)$ theory. The planar graph can be evaluated by using dimensional regularization \cite{Brandt:2001ud}
\beq
\Pi ^{\mu \nu} _{\rm planar}= 2e^2\int dx \int {d^d l \over (2\pi)^d} {2l^\mu l^\nu - g^{\mu \nu} l^2 -2x(1-x) q^\mu q^\nu + g^{\mu \nu}x(1-x)q^2 \over (l^2+ x(1-x)q^2)^2} 
\eeq
 and the result, as $d \rightarrow 2$, is 
\beq
\Pi ^{\mu \nu} _{\rm planar}= (g^{\mu \nu} -{q^\mu q^\nu \over q^2}){e^2 \over \pi} \,.
\eeq
The non-planar contribution is 
\bea
& & \Pi ^{\mu \nu} _{\rm non-planar}= \label{finite} \\
& & 2e^2\int dx \int {d^d l \over (2\pi)^d} {2l^\mu l^\nu - g^{\mu \nu} l^2 -2x(1-x) q^\mu q^\nu + g^{\mu \nu}x(1-x)q^2 \over (l^2+ x(1-x)q^2)^2} \exp il \theta q \,. \nonumber 
\eea
The exponent regularizes the UV divergences in the above expression \eqref{finite} and makes it finite. Indeed, by taking the trace of \eqref{finite} we obtain
\beq
g_{\mu \nu}\Pi ^{\mu \nu} _{\rm non-planar} \sim -2e^2 (d-2) \int {dt \over t^{d/2}} \exp (- {(\theta q)^2 \over t})
\eeq
and we observe that in the limit $d\rightarrow 2$ \eqref{finite} vanishes \cite{Armoni:2002fh}
\beq
 \Pi ^{\mu \nu} _{\rm non-planar}=0 \,. \label{zero2}
\eeq
And thus we learn that the vacuum polarization of the noncommutative theory differs from the vacuum polarization of the commutative theory. Whereas in the case of the commutative theory the total (planar plus non-planar) vacuum polarization is identically zero (since the adjoint fermion does not couple to the gauge field), it is non-zero in the noncommutative case. We observe that the limit $\theta \rightarrow 0$ is singular, since the commutative result is not recovered as we take the limit.

Let us now focus on generic planar graphs of the noncommutative theory. These graphs are selected in the large-$N$ limit of the $U(N)$ theory, as in the commutative case. While we cannot make general statements about generic values of $N$ and $\theta$, we would like to mention our result \eqref{zero2} that for any $N$ and any non-zero $\theta$ the one-loop non-planar graph vanishes. In particular, at the one-loop level, even the $U(1)$ noncommutative resembles the large-$N$ commutative theory, rather then the commutative $U(1)$ theory.

The large-$N$ commutative gauge theory with adjoint matter is not a solvable model. However, it is expected to confine and to admit infinitely many parallel Regge trajectories \cite{Kutasov:1993gq}, similarly to pure Yang-Mills theory in 4d. It is somewhat surprising that the noncommutative $U(1)$ 2d theory is expected to admit such a rich structure.

\section{Theories with fundamental matter}

In this section we consider the noncommutative $U(N)$ gauge theory with a single Dirac fermion in the fundamental representation. The action is

  \beq
S=\int d^2 x \, \left ( -{1\over 2e^2}{\rm tr}\, F_{\mu \nu} \star F ^{\mu \nu} + \bar \Psi i \slsh \partial \star \Psi + \bar \Psi \star \slsh A \star \Psi  \right )\, .
\eeq 

It is convenient to use the light-cone gauge $A_-=0$ and to work with light-cone coordinates, as used by 't Hooft in his seminal paper \cite{'tHooft:1974hx}. In this gauge the pure Yang-Mills
part of the action becomes free, $S_{\rm YM}=-\int d^2 x \, {1\over 2}{\rm tr}\, (\partial _- A_+)^2$. Therefore the only remnant of noncommutativity is in the gluon-fermion vertex. The Feynman rules of the theory are listed in fig.\eqref{rules} below. 

\begin{figure}[!ht]
\centerline{\includegraphics[width=4cm]{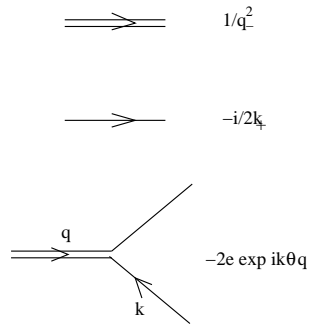}}
\caption{\footnotesize Feynman rules for noncommutative QCD in the light-cone gauge.} \label{rules}
\end{figure}

\begin{figure}[!ht]
\centerline{\includegraphics[width=4cm]{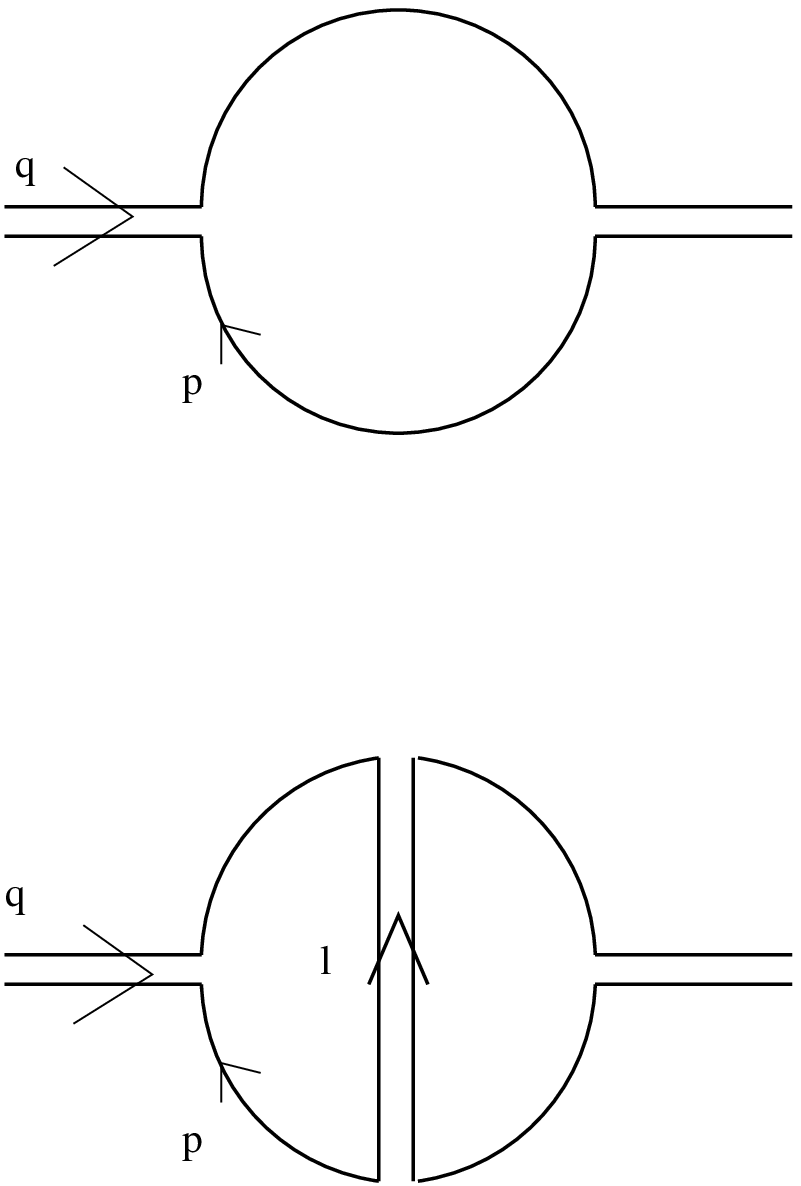}}
\caption{\footnotesize Planar one-loop and two-loop contributions to the vacuum polarization.} \label{graphs}
\end{figure}

Let us consider the vacuum polarization. The one-loop and two-loop contributions are planar, see figure \eqref{graphs}. It is therefore tempting to suggest that the mass spectrum of the commutative and noncommutative theories is identical \cite{Ardalan:2010qb}.

Starting from three-loop order, see figure \eqref{graph} below, there exists non-planar contributions which differ from their commutative counterparts. 

\begin{figure}[!ht]
\centerline{\includegraphics[width=4cm]{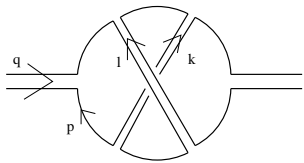}}
\caption{\footnotesize A non-planar three-loop contribution to the vacuum polarization.} \label{graph}
\end{figure}

Let us write down the expression for the three-loop non-planar diagram (for clarity we choose to write the expression in the Lorentz gauge)
\bea
& &  \Pi ^{\mu \nu} _{\rm non-planar} = e^6 \int {d^2 p \over (2\pi)^2} {d^2 l \over (2\pi)^2} {d^2 k \over (2\pi)^2} {1\over l^2} {1\over k^2} \exp (i 2k \theta l) \times \label{three} \\
& & {\rm tr}\, (\gamma ^\mu {1\over \slsh p + \slsh q} \gamma ^\rho {1\over \slsh p + \slsh q + \slsh l} \gamma ^\lambda {1\over \slsh p +\slsh q +\slsh l + \slsh k} \gamma ^\nu {1\over \slsh p + \slsh l +\slsh k } \gamma _\rho {1\over \slsh p + \slsh k} \gamma _\lambda {1\over \slsh p} ) \nonumber 
\eea 
When $\theta =0$ (the commutative theory), it is obvious from dimensional analysis, gauge invariance and Lorentz invariance that $\Pi _{\rm non-planar} =c_0 {e^6 \over q^6}$, where $c_0$ is a constant. It is then clear that when $\theta \ne 0$, the result of \eqref{three} must be
\beq 
\Pi _{\rm non-planar} =  {e^6 \over q^6} F(\theta q^2) \,. \label{expected}
\eeq
Since the commutative theory is super-renormalizable (see discussion in section 2) the above amplitude
admits a smooth expansion around $(\theta q^2)=0$, namely
\beq
F = c_0 + c_1 (\theta q^2) + c_2 (\theta q^2)^2 + ... \,.
\eeq 
While we did not calculate the three-loop diagram, it is manifestly $\theta$ dependent. Higher order non-planar graphs are also $\theta$ dependent. There is, therefore, no reason to believe that the mass spectrum of the theory should be identical to the mass spectrum of the commutative theory. Since the corrections arise from non-planar graphs, starting from three-loop order, a typical correction to the 't Hooft model meson masses should take in the large-$N$ and $e^2 \theta \ll 1$ limits, the form
\beq 
\delta M^2 \sim  (e^2N)(e^2 \theta)^2 \,.
\eeq     
We conclude by suggesting that due to the fact that for any $N$ there is qualitatively no difference between the commutative and the noncommutative theories (namely that the difference arise only at the three-loop order), even the $U(1)$ theory, namely the noncommutative version of the Schwinger model, could admit a spectrum similar to the spectrum of the large-$N$ 't Hooft model. This is, actually, not surprising, since it is well known that the noncommutative $U(1)$ theory is similar to the commutative $SU(N)$ theory \cite{Martin:1999aq}. Since the limit $\theta \rightarrow 0$ is smooth, the model can be simulated on the lattice. Preliminary results that support our conclusions are reported in ref. \cite{Bietenholz:2005iz}.

\section{Summary}

In this paper we discussed aspects of 2d noncommutative gauge field theories. There is a major difference between the 2d theory and the 4d theory: whereas the 4d noncommutative theory breaks explicitly Lorentz invariance, the 2d does not. As a result the analysis of the vacuum polarization of 2d gauge theories is much simpler with respect to the 4d analysis.    
 
We anticipated that in 2d UV/IR mixing are milder with respect to UV/IR mixing effects in 4d. This is indeed true in the case of two-dimensional QCD with fundamental fermions. In that case the one-loop and two-loop vacuum polarization graphs are planar and the first non-trivial effect appears at the three-loop order. A simple dimensional analysis reveals that the limit $\theta \rightarrow 0$ is smooth, contrary to the situation in higher dimensional theories.

We also considered two-dimensional QCD with adjoint fermions. In that case there is a non-planar one-loop contribution. It changes dramatically the behavior of the noncommutative theory and makes the limit $\theta \rightarrow 0$ singular.

In both cases we identified contributions that influence the mass spectrum of the theory. The effect is most dramatic for the $U(1)$ theory with an adjoint fermion: while the commutative theory is free, the noncommutative theory resembles the commutative four dimensional Yang-Mills gauge theory.
 
\newpage

{\it \bf Acknowledgements} I wish to thank Tim Hollowood and Prem Kumar for discussions. Special thanks to Neda Sadooghi for numerous discussions and comments on the manuscript.


\end{document}